\documentstyle[11pt,epsfig]{article}
\title{\bf Thermodynamics of the FRW universe at the event horizon in Palatini $f(R)$ gravity}
\author{A. S. Sefiedgar\thanks{e-mail: a.sefiedgar@umz.ac.ir} \,\,\, and \,\,
M. Mirzazadeh\thanks{email: mandana.mirzazadeh@yahoo.com}
\\ {\small Department of Physics, Faculty of Basic Sciences, University of Mazandaran, Babolsar 47416-95447, Iran}}
\begin{document}
\maketitle 
\begin{abstract}
In an accelerated expanding universe, one can expect the existence of an event horizon. It may be interesting to study the thermodynamics of the Friedmann-Robertson-Walker (FRW) universe at the event horizon. Considering the usual Hawking temperature, the first law of thermodynamics does not hold on the event horizon. To satisfy the first law of thermodynamics, it is necessary to redefine Hawking temperature. In this paper, using the redefinition of Hawking temperature and applying the first law of thermodynamics on the event horizon, the Friedmann equations are obtained in $f(R)$ gravity from the viewpoint of Palatini formalisn. In addition, the generalized second law (GSL) of thermodynamics, as a measure of the validity of the theory, is investigated. 
\end{abstract}
\vspace{2cm}
\section{Introduction}
The existence of a deep connection between gravity and thermodynamics is one of the greatest discoveries in theoretical physics \cite{1,2,3,4,5}.
Based on the Hawking proposal, a schwarzschild black hole emits a thermal radiation. For a black hole with mass $M$, the Hawking temperature is given by $T=\frac{1}{8\pi M}$ \cite{4,5}. Moreover, the black hole entropy is given by $S=\frac{A}{4l_p^2}$, which is introduced by Bekenstein. The parameter $A$ is the area of the black hole horizon and $l_p$ is the Planck length. The black hole mass, temperature and the horizon entropy obey the first law of thermodynamics. Hence, one can conclude that a black hole can be assumed as a thermodynamical object \cite{7,8,9}.
In 1995, Jacobson has considered the relation between the horizon area and the entropy and derived the Einstein field equations from the first law of thermodynamics \cite{jacobson}. Following Jacobson, the relation between the thermodynamics and the field equations has been investigated in modified theories of gravity \cite{11,12,13,14,15,16}. 
The thermodynamical interpretation of gravity is also important in modern cosmology. It is possible to derive the Friedmann equations from the first law of thermodynamics and vice versa \cite{11+,12+,13+}. The deep connection between the first law of thermodynamics and Friedmann equations has been investigated in gravity with Gauss-Bonnet term \cite{18}, Lovelock gravity theory \cite{18}, the braneworld scenarios \cite{19}, rainbow gravity \cite{daghigh}, scalar-tensor gravity and $f(R)$ gravity \cite{20}.

The thermodynamics of the FRW universe bounded by the apparent horizon has been studied by many authors \cite{11+,12+,13+}. However, astronomical observations show that the universe is currently in an accelerated expanding phase \cite{14+}. In an accelerated expanding universe, there may exist an event horizon \cite{wang,chen}. Certainly, it is necessary to investigate the thermodynamics on the event horizon.
Through the researches about the thermodynamics of the universe, Wang et al. have found that the laws of thermodynamics do not hold on the event horizon. Therefore, one may conclude that the event horizon is unphysical \cite{wang}. Chacraborty has proposed a redefinition of Hawking temperature to obtain the universe bounded by the event horizon as a Bekenstein system \cite{chakraborty}. Using the Chakraborty redefinition of Hawking temperature, Tu and Chen have considered a universe dominated by tachyon fluid and obtained a good thermodynamical description on the event horizon \cite{Tu}. However, the temperature redefinition proposed by Chakraborty can be applied only in a flat universe. Tu and Chen have introduced a new redefinition of the temperature in a universe with an arbitrary spatial curvature \cite{chen}. They have investigated the thermodynamics on the event horizon in metric $f(R)$ gravity. Utilizing the redefinition of the temperature, it is possible to study the thermodynamics of the event horizon in Palatini $f(R)$ gravity.

In this paper, we are going to apply Hawking temperature redefinition introduced by Tu and Chen in \cite{chen} to investigate the event horizon thermodynamics from the viewpoint of Palatini $f(R)$ gravity.
Redefining the gravitational energy density and the gravitational pressure density, the continuity equation has been satisfied. Then, we have considered the new gravitational source accompanying with the ordinary matter as a modified energy-momentum source in the context of general relativity. 
Applying the first law of thermodynamics on the event horizon and using the usual entropy-area relation, we have derived the Friedmann equations the same as the ones obtained via other approaches. 
Since the first law of thermodynamics is reasonably applicable on the event horizon, one can consider the event horizon as a physical system with an equilibrium thermodynamics in Palatini $f(R)$ gravity.
Moreover, the equivalency between the first law of thermodynamics and Friedmann equations has been appeared.
Finally, the generalized second law of thermodynamics (GSL) has been studied. The generalized second law of thermodynamics can be satisfied by choosing suitable $f(R)$ functions.
\section{Hawking temperature redefinition on the event horizon}
Considering a homogeneous and isotropic FRW universe, one can write the metric as follows
\begin{equation}
ds^2=h_{ij}dx^idx^j+R^2d\Omega_2^2,
\end{equation}
where  $h_{ij}$ is the two-dimensional metric
\begin{equation}
h_{ij}=diag\left(-1,\frac{a^2(t)}{1-Kr^2}\right).
\end{equation}
The parameters $i$ and $j$ take the values $0$ and $1$. The parameter $K$ is the spatial curvature constant, the parameter $R=a(t)r$ is the area radius and $a(t)$ is the scale factor. Using $R$ as a scalar field in the normal 2-dimensional space, one can define a scalar quantity 
\begin{equation}
\chi=h^{ij}\partial_i R \partial_j R.
\end{equation} 
The apparent horizon can be found via $\chi=0$ as
\begin{equation}
R_A=\frac{1}{\sqrt{H^2+\frac{K}{a^2}}},
\end{equation}
where $H=\frac{1}{a}\frac{da}{dt}=\frac{\dot{a}}{a}$ is the Hubble parameter. 

The surface gravity on the apparent horizon is given by
\begin{equation}
\kappa_A=-\frac{1}{2}\frac{\partial \chi}{\partial R}|_{R=R_A}=\frac{1}{R_A}.
\end{equation}
Then, one can derive Hawking temperature on the apparent horizon as
\begin{equation}
T_A=\frac{|\kappa_A|}{2\pi}=\frac{1}{2\pi R_A}.
\end{equation}
On the other hand, one can consider the definition of the event horizon
\begin{equation}\label{RE1}
R_E=a(t)\int_t^{\infty}\frac{dt}{a(t)}.
\end{equation}
In an accelerated expanding universe, one can find that the infinite integral converges. In other words, it is possible to have a cosmological event horizon in an accelerated expanding universe. Then, the thermodynamics on the event horizon is important to be studied. 

Through the researches about the thermodynamics on the event horizon, wang et al. found that the thermodynamical description based on the standard definitions of boundary entropy and temperature breaks down in a universe bounded by the cosmological event horizon \cite{wang}. They concluded that the cosmological event horizon is unphysical from the viewpoint of thermodynamic laws. However, a redefinition of Hawking temperature has been introduced by Chakraborty in a flat universe to preserve the laws of thermodynamics on the event horizon \cite{chakraborty}. The temperature redefinition has been upgraded by Tu et al. to be applicable in a universe with an arbitrary curvature constant \cite{chen}. According to \cite{chen}, one can write the surface gravity on the event horizon as
\begin{equation}
\kappa_E=-\frac{1}{2}\frac{\partial \chi}{\partial R}|_{R=R_E}\frac{\dot{R}_A}{R_A}\frac{R_E}{\dot{R}E}.
\end{equation}   
Using the relation between the surface gravity and Hawking temperature on spacetime  horizons, one can find
\begin{equation}\label{TE1}
T_E=\frac{|\kappa_E|}{2\pi}=\frac{H}{2\pi}\left(\frac{K}{a^2}-\dot{H}\right)\frac{R^2_E}{\dot{R}_E}.
\end{equation}

To investigate the universality of the redefinition of Hawking temperature on the event horizon, one can investigate the validity of the first law of thermodynamics. During an infinitesimal time interval, one can write the energy flux across the event horizon as \cite{wang,chen,chakraborty,24-,25-}
\begin{equation}\label{deltaQ0}
\delta Q=A T_{\mu \nu} k^{\mu} k^{\nu} dt|_{r=R_E},
\end{equation}
where $k^{\mu}$ is a null vector, $T_{\mu \nu}=(\rho+p)u_{\mu}u_{\nu}+pg_{\mu\nu}$ is the perfect fluid energy-momentum tensor and $u^{\mu}$ is the timelike 4-velocity vector. Thus, the energy flux can be obtained as 
\begin{equation}
\delta Q=4 \pi R_E^3H(\rho +p)dt.
\end{equation}
It can also be written as 
\begin{equation}\label{deltaQ}
\delta Q= \frac{H R_E^3}{G}\left(\frac{K}{a^2}-\dot{H}\right)dt,
\end{equation}
in which the Friedmann equation in Einstein gravity, $\dot{H}-\frac{K}{a^2}=-4\pi G(\rho+p)$ has been used.
On the other hand, the Bekenstein entropy-area relation in Einstein gravity is given by
\begin{equation}\label{entropy}
S_E=\frac{A}{4G}.
\end{equation}
Using equations (\ref{TE1}) and (\ref{entropy}), one can  find
\begin{equation}\label{TdS}
T_E dS_E=\frac{HR_E^3}{G}\left(\frac{K}{a^2}-\dot{H}\right)dt.
\end{equation}
Comparing equation (\ref{deltaQ}) and (\ref{TdS}), the validity of the first law of thermodynamics, $\delta Q=T_E dS_E$, can be concluded on the event horizon \cite{chen}.
It seems that the temperature redefinition can lead to a good thermodynamical description on the event horizon.

\section{From the first law of thermodynamics into Friedmann equations in Palatini $f(R)$ gravity}
The action of $f(R)$ gravity in Palatini formalism is written as
\begin{equation}
I=\int d^4x \sqrt{-g}\left[\frac{f({\mathcal{R}})}{2{\mathit{k}}^2}+ {\mathcal{L}}_{matter}\right],
\end{equation}
where $g$ is the determinant of the metric tensor, ${\mathcal{L}}_{matter}$ is the matter lagrangian and ${\mathit{k}}^2=8\pi G$. The connection and the metric tensor are considered as independent variables in Palatini formalism.
The Ricci tensor and the Ricci scalar constructed with the independent connection  are denoted respectively by ${\mathcal{R}}_{\mu \nu}$ and ${\mathcal{R}}=g^{\mu \nu}{\mathcal{R}}_{\mu \nu}$. Certainly, ${\mathcal{R}}_{\mu \nu}$ is different from the Ricci tensor $R_{\mu \nu}$ which is constructed with the Levi-Civita connection of the metric.  Variation of the action with respect to the metric leads to
\begin{equation}\label{field1}
F({\mathcal{R}}){{\mathcal{R}}}_{\mu \nu}-\frac{1}{2}f({\mathcal{R}})g_{\mu \nu}={\mathit{k}}^2T_{\mu \nu}^{m},
\end{equation}
in which $F({\mathcal{R}}) \equiv df({\mathcal{R}})/d {\mathcal{R}}$ and $T_{\mu \nu}^{m}$ is the energy-momentum tensor related to the ordinary matter. 
Varying the action with respect to the connection and using equation (\ref{field1}) yield
\begin{equation}
FG_{\mu \nu}={\mathit{k}}^2 T_{\mu \nu}^{m}-\frac{1}{2}g_{\mu \nu} (F{\mathcal{R}}-f)+{\nabla}_{\mu}{\nabla}_{\nu}F-g_{\mu \nu}g^{\alpha \beta}{\nabla}_{\alpha}{\nabla}_{\beta}F-\frac{3}{2F}\left[{\nabla}_{\mu}F{\nabla}_{\nu}F-\frac{1}{2}g_{\mu \nu}g^{\alpha \beta}{\nabla}_{\alpha}F{\nabla}_{\beta}F \right].
\end{equation}
Here $G_{\mu \nu}=R_{\mu \nu}-\frac{1}{2}Rg_{\mu \nu}$ is the Einstein tensor and $T_{\mu \nu}^{m}=(\rho_m+p_m)u_{\mu}u_{\nu}+p_mg_{\mu \nu}$ is the energy-momentum tensor of the ordinary matter. One can rewrite the field equations as
\begin{equation}
G_{\mu \nu}=\frac{{\mathit{k}}^2}{F}\left({T_{\mu \nu}^{m}}+{T_{\mu \nu}^{g}} \right),
\end{equation}
where
\begin{equation}
T_{\mu \nu}^g=\frac{1}{{{\mathit{k}}^2}} \left( \frac{f-F{\mathcal{R}}}{2}g_{\mu \nu}+{\nabla}_{\mu}{\nabla}_{\nu}F-g_{\mu \nu}g^{\alpha \beta}{\nabla}_{\alpha}{\nabla}_{\beta}F\right)-\frac{3}{2F{{\mathit{k}}^2}}\left[{\nabla}_{\mu}F{\nabla}_{\nu}F-\frac{1}{2}g_{\mu \nu}g^{\alpha \beta}{\nabla}_{\alpha}F{\nabla}_{\beta}F \right]   ,
\end{equation}
is the gravitational energy-momentum tensor arising from $f(R)$ gravity.
Now, the gravitational energy density and the gravitational pressure can be written respectively as
\begin{equation}\label{rhog}
\rho_g=\frac{1}{{{\mathit{k}}^2}}\left[\frac{1}{2}(F{\mathcal{R}}-f)-3H\dot{F}-\frac{3}{4}\frac{{\dot{F}}^2}{F}  \right],
\end{equation}
and 
\begin{equation}\label{pg}
p_g=\frac{1}{{{\mathit{k}}^2}}\left[-\frac{1}{2}(F{\mathcal{R}}-f)+\ddot{F}+2H\dot{F}-\frac{3}{4}\frac{{\dot{F}}^2}{F}  \right].
\end{equation}

The perfect fluid energy-momentum tensor satisfies the continuity equation
\begin{equation}\label{perfect}
\dot{\rho}_m+3H(\rho_m+p_m)=0.
\end{equation}
It is easy to investigate the continuity equation for the gravitational sector of the energy-momentum tensor arising from $f(R)$ gravity. Clearly, one can find
\begin{equation}
\dot{\rho}_g+3H(\rho_g+p_g)=\frac{3}{{{{\mathit{k}}^2}}}\left(H^2+\frac{K}{a^2}\right)\dot{F}.
\end{equation}
To satisfy the continuity equation, it is possible to redefine the gravitational energy density and the gravitational pressure density as 
\begin{equation}\label{rhog+}
\hat{\rho}_g=\frac{1}{{{\mathit{k}}^2}}\left[\frac{1}{2}(F{\mathcal{R}}-f)-3H\dot{F}-\frac{3}{4}\frac{{\dot{F}}^2}{F}+3(1-F)\left(H^2+\frac{K}{a^2}\right) \right],
\end{equation}
and
\begin{equation}\label{pg+}
\hat{p}_g=\frac{1}{{{\mathit{k}}^2}}\left[-\frac{1}{2}(F{\mathcal{R}}-f)+\ddot{F}+2H\dot{F}-\frac{3}{4}\frac{{\dot{F}}^2}{F} -(1-F)\left(2\dot{H}+3H^2+\frac{K}{a^2}\right) \right].
\end{equation}
Now, one can rewrite the continuity equation for the gravitational energy density and gravitational pressure density as
\begin{equation}\label{hat}
\dot{\hat{\rho}}_g+3H(\hat{\rho}_g+\hat{p}_g)=0.
\end{equation}
From equations (\ref{perfect}) and (\ref{hat}), one can write the total continuity equation as

\begin{equation}\label{contiT}
\dot{\rho}_T+3H(\rho_T+p_T)=0,
\end{equation}
in which $\rho_T=\rho_m+\hat{\rho}_g$ and $p_T=p_m+\hat{p}_g$.
It is clear that the redefinition of the energy density in equation (\ref{rhog+}) and the pressure density in equation (\ref{pg+}) lead to the standard continuity equation. It is now possible to consider $\rho_T$ and $p_T$ as the energy and the pressure density of a modified source in the context of general relativity. Hence, one can apply the usual entropy-area relation in (\ref{entropy}).

To derive the Friedmann equation, one can start with the first law of thermodynamics. Using equation (\ref{deltaQ0}), the energy flux can be written as
\begin{equation}\label{dQ+}
\delta Q=4 \pi R_E^3H({\rho}_T+{p}_T)dt.
\end{equation} 
Utilizing equations (\ref{TdS}) and (\ref{dQ+}), the first law of thermodynamics yields
\begin{equation}\label{TdS1}
\left(\frac{K}{a^2}-\dot{H} \right)=4\pi G (\rho_T+{p}_T).
\end{equation}
Now, one can Substitute equations (\ref{rhog+}) and (\ref{pg+}) into equation (\ref{TdS1}) to find the Friedmann equation
\begin{equation}\label{Fried1}
-2 \left( \dot{H} -\frac{K}{a^2} \right)F={{\mathit{k}}^2}(\rho_m+p_m)+\ddot{F}-H\dot{F}-\frac{3}{2}\frac{\dot{F}^2}{F}.
\end{equation} 
The other Friedmann equation can be obtained via equations (\ref{contiT}) and (\ref{Fried1})
\begin{equation}\label{Fried2}
3\left( H^2+\frac{K}{a^2}   \right)F={{\mathit{k}}^2}\rho_m+\frac{1}{2}(F{\mathcal{R}}-f)-3H\dot{F}-\frac{3}{4}\frac{\dot{F}^2}{F}.
\end{equation}
The modified Friedmann equations in (\ref{Fried1}) and (\ref{Fried2}) are the same as the ones obtained in \cite{palatini}. In other words, the first law of thermodynamics on the event horizon leads to the Friedmann equations which are consistent with the ones obtained via the other approach in \cite{palatini}. Hence, one can rely on the validity of the first law of thermodynamics on the event horizon and consider the event horizon as a thermodynamical system in Palatini $f(R)$ gravity. It is also clear that the first law of thermodynamics is equivalent with Friedmann equations.

Equation (\ref{Fried2}) can also be rewritten as
\begin{equation}
\left( H^2+\frac{K}{a^2}   \right)=\frac{8\pi G}{3}\rho_{eff},
\end{equation}
where
\begin{equation}
\rho_{eff}=\frac{1}{{F}}\left[{\rho_m}+\frac{1}{{8\pi G}}\left(\frac{1}{2}(F{\mathcal{R}}-f)-3H\dot{F}-\frac{3}{4}\frac{\dot{F}^2}{F}\right)\right],
\end{equation}
can be considered as an effective energy density. Such an effective energy density may provide an accelerated expanding universe to solve the problem of dark energy.
The modified Friedmann equations can be reduced to the Friedmann equations in Einstein gravity by putting $f(\mathcal{R})=\mathcal{R}$.
 
\section{The generalized second law of thermodynamics}
Applying the usual Hawking temperature, the first law of thermodynamics does not hold on the event horizon. However, using the redefinition of Hawking temperature, it has been shown that the first law of thermodynamics may hold on the event horizon in Einstein gravity and metric $f(R)$ gravity \cite{chen}. Hence, the FRW universe bounded by the event horizon may be described by equilibrium thermodynamics. In previous section, we have considered Hawking temperature redefinition on the event horizon in Palatini $f(R)$ gravity. We have shown that the first law of thermodynamics can yield the Friedmann equations just the same as the ones obtained via the other approach in \cite{palatini}. It means that he FRW universe bounded by the event horizon can be described by an equilibrium thermodynamics in Palatini approach too. 

It is now possible to investigate the generalized second law of thermodynamics in the FRW universe bounded by the event horizon in Palatini approach. Considering a universe with the dust energy-momentum tensor in $f(R)$ gravity, one can write the continuity equations as
\begin{equation}
\dot{\rho_d}+3H\rho_d=0,
\end{equation}
and
\begin{equation}
\dot{\hat{\rho}}_g+3H(\hat{\rho}_g+\hat{p}_g)=0.
\end{equation}
The parameter $\rho_d$ is the energy density of dust and the pressure of dust is $p_d=0$.
The energy flux can be obtained from equation (\ref{deltaQ0}) as
\begin{equation}
\delta Q=4\pi R_E^3H(\rho_d+\hat{\rho}_g+\hat{p}_g)dt.
\end{equation}
Based on the generalized second law of thermodynamics, the sum of the entropy on the event horizon and the one inside the bulk must never decrease. To find the total entropy changes, one can start with Gibb's relation \cite{33-,34-,35-}
\begin{equation}\label{gibb1}
T_EdS_{in}=dE_{in}+\hat{p}_gdV,
\end{equation}
where $V=\frac{4 \pi R_E^3}{3}$. The parameters $E_{in}=\frac{4 \pi R_E^3}{3}(\rho_d+\hat{\rho}_g)$ and $S_{in}$ are the energy and the entropy inside the boundary respectively. 
From equation (\ref{RE1}), one can find
\begin{equation}\label{dRE1}
dR_E=(HR_E-1)dt.
\end{equation}
Now, equation (\ref{gibb1}) leads to
\begin{equation}
\frac{dS_{in}}{dt}=-\frac{4 \pi R_E^2}{T_E}(\rho_d+\hat{\rho}_g+\hat{p}_g),
\end{equation}
in which we have used the relation (\ref{dRE1}). Since the universe is assumed to be in thermal equilibrium, the temperature inside the universe and the temperature on the event horizon are considered the same.  
Using the definition $S_{tot}=S_{in}+S_E$ and applying equation (\ref{TdS}), one can find
\begin{equation}\label{ssdot}
\frac{dS_{tot}}{dt}=\frac{4 \pi R_E^2}{T_E}(\rho_d+\hat{\rho}_g+\hat{p}_g)(HR_E-1).
\end{equation}
It is possible to substitute equations (\ref{rhog+}) and (\ref{pg+}) into equation (\ref{ssdot}) to find
$$\frac{dS_{tot}}{dt}=\frac{ R_E^2}{2G T_E}(HR_E-1)\left(\rho_d+\ddot{F}-H\dot{F}-\frac{3}{2}\frac{\dot{F}^2}{F}\right)-$$
\begin{equation}
\frac{R_E^2}{GT_E}(1-F)(HR_E-1)\left(\dot{H}-\frac{K}{a^2}\right).
\end{equation}
Based on the redefinition of Hawking temperature, we have found the event horizon as an equilibrium thermodynamical system. Hence, the first law of thermodynamics and the Gibb's relation have been applied to obtain the change of the total entropy. To satisfy GSL, it is necessary to have $\dot{S}_{tot}\geq 0$.
Of course, the validity of the generalized second law of thermodynamics depends on the $f(R)$ functions. In other words, GSL can provide some constraints on the choice of $f(R)$ gravity models. 

To investigate GSL, one can substitute $T_E$ from equation (\ref{TE1}) into equation (\ref{ssdot}) to find 
\begin{equation}
\frac{dS_{tot}}{dt}=\frac{8 \pi^2 }{H\left(\frac{K}{a^2}-\dot{H}\right)}(\rho_d+\hat{\rho}_g+\hat{p}_g)\dot{R}_E^2.
\end{equation}
Based on the equation ({\ref{TdS1}), one can conclude that $\left(\frac{K}{a^2}-\dot{H}\right) \geq 0$ in the weak energy condition. Hence, GSL can be satisfied, when
\begin{equation}\label{cond1}
(\rho_d+\hat{\rho}_g+\hat{p}_g) \geq 0.
\end{equation}  
It means that the function $f$ should satisfy the condition
\begin{equation}\label{cond2}
\rho_d+\ddot{F}-H\dot{F}-\frac{3}{2}\frac{\dot{F}^2}{F^2}+2(1-F)\left(\frac{K}{a^2}-\dot{H}\right) \geq 0.
\end{equation} 
In the case $f(R)=R$, one can substitute $F=1$ in equation (\ref{cond2}) to investigate the validity of GSL in general relativity.  Since $ \rho_d \geq 0$, one can find $\dot{S}_{tot} \geq 0$  in the dust dominated universe bounded by the event horizon in general relativity.  The Hawking temperature redefinition has provided the validity of the GSL in the universe bounded by the event horizon in general relativity.

\section{Conclusions}
Although the apparent horizon thermodynamics has been studied by many authors in FRW universe, the event horizon thermodynamics is not investigated enough. Since the universe is undergoing an accelerated expansion phase, one may expect the existence of an event horizon. Certainly, the thermodynamics of the event horizon is important to be studied. It has been shown that applying the usual Hawking temperature leads to a non-equilibrium thermodynamics on the event horizon \cite{wang}. Hence, the event horizon may be considered unphysical from the view point of the thermodynamic laws. However, redefining Hawking temperature can solve this problem \cite{chen,chakraborty}. The Hawking temperature redefinition introduced by Tu and Chen \cite{chen} leads to an equilibrium thermodynamics on the event horizon in general relativity and metric $f(R)$ gravity. In this paper, the thermodynamics of the event horizon in FRW universe is investigated in Palatini $f(R)$ gravity. Apparently, the continuity equation does not hold in Palatini formalism. To satisfy the continuity equation, redefinitions of the energy density and the pressure density have been introduced in the gravitational sector of the energy-momentum tensor. The redefined gravitational source accompanying with the ordinary  matter may play the role of a new modified source in the context of Einstein gravity. Considering this modified energy-momentum source, using the first law of thermodynamics on the event horizon and applying Bekenstein-Hawking entropy-area relation, the Friedmann equations have been derived. The corrections to the Friedmann equations are the same as the ones obtained via other approaches in \cite{palatini}. It means that considering the first law of thermodynamics on the event horizon is reliable. In addition, the equivalency of the first law of thermodynamics and the Friedmann equations has been shown. The generalized second law of thermodynamics as a measure of the validity of the gravitational models are investigated in Palatini $f(R)$ gravity. Of course, GSL can provide some constraints on choosing $f(R)$ functions.

\end{document}